\documentclass[twocolumn,floatfix,showpacs,prb,superscriptaddress]{revtex4}%

\usepackage{epsfig}
\usepackage{dcolumn}
\usepackage{bm}
\usepackage{amsmath}
\usepackage{graphicx}
\usepackage{amssymb}

\begin{document}

\title{Phase-Locking Transition in a Chirped Superconducting Josephson Resonator}

\author{O. Naaman}
\affiliation{Quantum Nanoelectronics Laboratory, Department of Physics, University of California, Berkeley, California 94720, USA}
\author{J. Aumentado}
\affiliation{National Institute of Standards and Technology, 325 Broadway, Boulder, Colorado 80305, USA}
\author{L. Friedland}
\affiliation{Racah Institute of Physics, Hebrew University, Jerusalem 91904, Israel}
\author{J. S. Wurtele}
\affiliation{Department of Physics, University of California, Berkeley, California 94720, USA}
\affiliation{Center for Beam Physics, Lawrence Berkeley National Laboratory, Berkeley CA 94720, USA}
\author{I. Siddiqi}
\affiliation{Quantum Nanoelectronics Laboratory, Department of Physics, University of California, Berkeley, California 94720, USA}


\date{May 14, 2008}




\pacs{85.25.Cp, 05.45.-a, 03.67.Lx, 74.50.+r}
\maketitle

{\bf By coupling a harmonic oscillator to a quantum system it is possible to perform a dispersive measurement \cite{Kimble98,Grangier98} that is quantum non-demolition (QND), with minimal backaction \cite{Braginsky}. A \emph{non-linear} oscillator has the advantage of measurement gain, but what is the backaction? Experiments on superconducting quantum bits (qubits) coupled to a non-linear Josephson oscillator \cite{Siddiqi06,Lupascu06,Boulant07} have thus far utilized the switching of the oscillator near a dynamical bifurcation for sensitivity\cite{Siddiqi04,Siddiqi05}, and have demonstrated partial QND measurement.  The detailed backaction associated with the switching process is complex, and may ultimately limit the degree to which such a measurement can be QND. Here we demonstrate a new dynamical effect in Josephson oscillators by which the bifurcation can be accessed without switching. When energized with a frequency chirped drive with an amplitude close to a sharp phase-locking threshold, the oscillator evolves smoothly in one of two diverging trajectories \cite{Fajans99}\textemdash a pointer for the state of a qubit. The observed critical behavior agrees well with theory and suggests a new modality for quantum state measurement.}

A superconductor-insulator-superconductor tunnel junction is a unique electrical circuit element that can be simultaneously non-linear and non-dissipative. The Josephson equations parameterize the non-linear tunnel current $I=I_0\sin\delta$ and voltage $V_\textrm{J}=\hbar\dot\delta/2e$ in terms of the gauge-invariant superconducting phase difference $\delta$ across the junction and its time derivative $\dot\delta$. These equations describe a non-linear inductor with inductance $L_\textrm{J}(\delta)=\hbar/(2eI_0\cos\delta)$, which can be shunted with a low-loss reactance \cite{Vlad07} to form a high quality-factor ($Q$) anharmonic oscillator. Coupling this Josephson oscillator to a qubit produces a state dependent shift of the resonant frequency\textemdash thus realizing a dispersive measurement \cite{Ilichev03,Lupascu04,Wallraff04}. When probed with a small number of photons, the Josephson oscillator is essentially harmonic and the measurement does not project the quantum state out of the qubit basis: it is therefore quantum non-demolition (QND), and has a well characterized minimal backaction \cite{Kimble98, Grangier98,Blais04,Schuster05}. When driven more strongly, anharmonicity causes the resonant frequency to vary with the amplitude of oscillation \cite{LandauLifshitz}, enhancing the oscillator's driven response to a frequency shift and thus ``amplifying" the qubit ``signal". Since the junction has vanishing internal dissipation, this process may also result in minimal decoherence. Moreover, the Josephson oscillator can also bifurcate \cite{Siddiqi05} into two metastable oscillation states\textemdash one which is low-amplitude and phase lagging and one which is high-amplitude and phase leading\textemdash resulting in a projective measurement. This has been demonstrated in the Josephson bifurcation amplifier (JBA) \cite{Siddiqi04} in which the oscillator is probed with a fixed frequency drive with varying amplitude \cite{Siddiqi06,Lupascu06}. The dynamical state of the oscillator depends on the qubit state, and measurement sensitivity arises from a sharp threshold for switching between the two dynamical states. While such a non-linear oscillator has the advantage of measurement gain, the precise nature of its backaction on the quantum system, especially that associated with the switching process, and the degree to which a QND measurement is possible \cite{Boulant07} is still being investigated. 

We demonstrate a new measurement scheme that still drives the Josephson oscillator to the bifurcation regime but does not involve any switching process. Instead of applying an amplitude modulated drive, we apply a chirped, microwave frequency drive to access a phenomenon known as autoresonance \cite{Fajans01}. In response to a frequency modulation, the oscillator may either phase-lock to the chirped carrier and latch to the high-amplitude oscillation state, or not lock to the drive and remain in the small-oscillation state. These two outcomes are separated in parameter space by a sharp threshold \cite{Fajans99b} that scales with the chirp rate and is sensitive to the junction $I_0$. Throughout its evolution, the system tracks a single basin of attraction\textemdash no switching occurs, thus avoiding any potential backaction associated with switching dynamics and transient oscillations. We call this new measurement device a Josephson chirped amplifier (JCA). We show that the observed threshold behavior is in excellent agreement with both analytic theory and numerical simulations, and estimate the performance of the JCA as a quantum state readout device.

\begin{figure}[t]
\epsfxsize=3.2in
\epsfbox{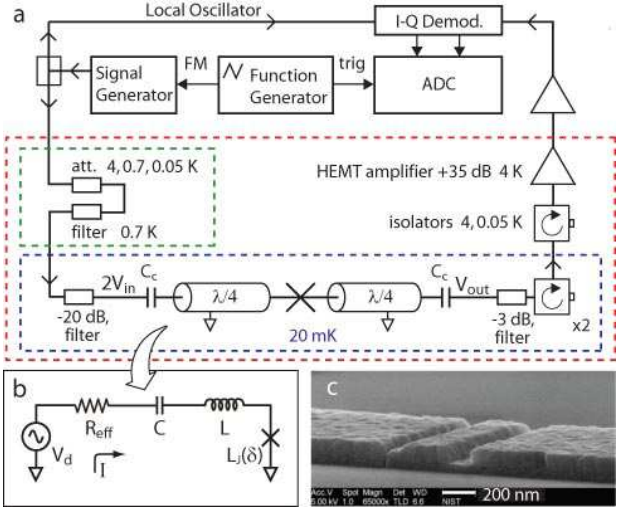}
\caption{\label{circuit} {\bf Experimental setup and circuit model.} {\bf a,} Schematic of the experimental setup. {\bf b,} Series $RLC$ model of the resonator. {\bf c,} Scanning electron micrograph of the junction.}
\end{figure}

In our experiment, the Josephson oscillator was formed by an Al/AlO$_x$/Al Josephson tunnel junction (Fig.\ \ref{circuit}c) placed in the middle of a high-$Q$ Nb half-wave coplanar waveguide resonator with a characteristic impedance $Z_0=50\;\Omega$, and symmetrically coupled to the 50 $\Omega$ environment via capacitors $C_c\approx10$ fF. The measurements were performed in a dilution refrigerator at $T$=20 mK; the experimental setup is shown schematically in Fig.\ \ref{circuit}a. Microwave excitation was applied to the resonator using an HP8780A vector signal generator, frequency-modulated with a triangle waveform to provide a linear, phase-continuous frequency chirp with a 50 MHz span. Typical chirp rates, $\alpha=d\omega/dt$, ranged from $\alpha/2\pi=10^{11}$ Hz/s to $10^{13}$ Hz/s. The transmitted microwave signal, $V_\textrm{out}$, was amplified and then demodulated (1.9 MHz IF bandwidth) to find its amplitude and phase. The amplitude of current oscillations in the resonator is given by $I=(2|V_\textrm{out}|/Z_0)\times\sqrt{Q/\pi}$.

\begin{figure}
\epsfxsize=3.2in
\epsfbox{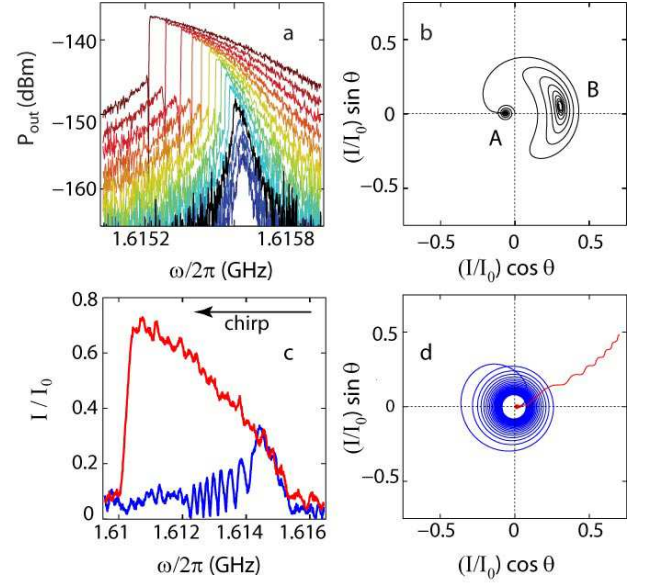}
\caption{\label{phase} {\bf Microwave transmission and phase space trajectories.} {\bf a,} Transmitted power $P_\textrm{out}$ in steady state, measured with a network analyzer. Input powers are $-155.5$~dBm (blue) to $-123$~dBm at 2.5 dB step. The system exhibits bifurcation above a critical power $P_c=-148$ dBm ($1.58\times10^{-18}$ W). {\bf b,} Simulated in-phase and quadrature components of current oscillations for detuning $(\omega-\omega_0)/2\pi=0.5$ MHz and drive power ramped in time from -126 dBm to -120 dBm. A and B label the low- and high-amplitude attractors, $\theta$ is the phase of the oscillations. {\bf c,} Response of the resonator to a chirped drive, $\alpha/2\pi=5\times10^{11}$ Hz/s, at $P_\textrm{in}=-126$ dBm (blue) and $-120$~dBm (red). {\bf d,} Simulated quadratures of the current in a chirped resonator with the same parameters as the data in c. The simulation was terminated at $\omega/2\pi=1.6106$ GHz.}

\end{figure}

We first measure the oscillator in steady state (Fig.\ \ref{phase}a). From the microwave transmission, $P_\textrm{out}$, at low power in the linear regime (Fig.\ \ref{phase}a, blue curve), we measure $\omega_0/2\pi=1.61564$ GHz and infer a quality factor $Q=27500\pm1000$. The $Q$ is limited by coupling to the $50\;\Omega$ environment and is within 10\% of its predicted value $Q=\pi/4Z_0^2\omega_0^2C_c^2$. When the drive power exceeds a critical value $P_{c}=-148$ dBm, the oscillator response bifurcates into two branches \cite{Siddiqi05}. From the measured $P_{c}$ we estimate the junction critical current $I_0=0.61\pm0.04\;\mu$A; this value is consistent with room temperature dc resistance measurements on co-fabricated junctions. The stated uncertainties are dominated by possible cryogenic variations in the attenuation of the coaxial lines attached to the resonator. Additionally, the observed power at which bifurcation occurs at different drive frequencies agrees well with theory \cite{Vlad07} using the measured values of $P_{c}$ and $Q$. 

\begin{figure}
\epsfxsize=3.2in
\epsfbox{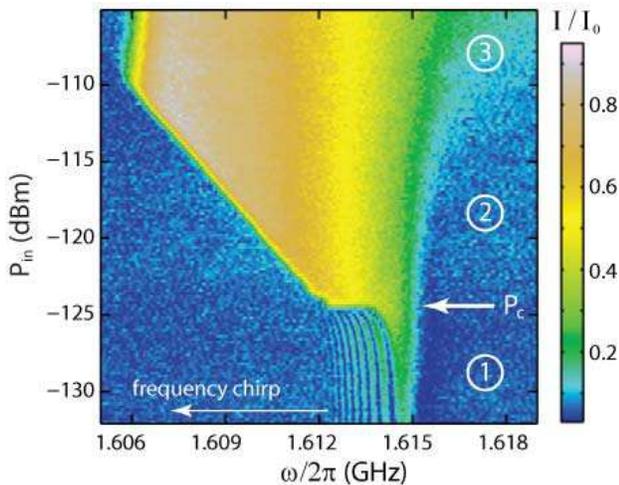}
\caption{\label{AR} {\bf Microwave response of a chirped non-linear resonator.} Normalized amplitude of current oscillations in the resonator as a function of drive power and frequency, with the frequency swept downwards at a rate of $5\times10^{11}$ Hz/s. The arrow indicates the critical power $P_c$ for phase-locking. In region 1 where $P_\textrm{in}$ is sub-critical, the oscillation amplitude grows in the vicinity of $\omega_{0}$ and rings back to zero below $\omega_{0}$. Above $P_c$, the oscillator phase-locks and its amplitude grows until it reaches a maximum set either by damping (region 2) or by $I_{0}$ (region 3).}   
\end{figure}

Having characterized its parameters, we proceed to excite the oscillator with a chirped microwave drive. Fig.\ \ref{phase}c shows the resonator response to a downwards frequency chirp at the rate of $\alpha/2\pi=5\times10^{11}$ Hz/s. At low power (blue) the amplitude of oscillations initially increases as the chirp passes through $\omega_0$. However, as the chirp progresses towards lower frequencies, the resonator decouples from the drive and rings down to rest. For a stronger drive amplitude (red) the response of the resonator changes dramatically: as the chirp passes through $\omega_0$, the resonator phase becomes locked to the drive and its amplitude grows with time. A threshold for phase-locking can be seen in Fig.\ \ref{AR}, which shows the normalized amplitude of current oscillations, $I/I_0$, as a function of frequency and drive power for a fixed chirp rate. In region 1, at low powers, no phase locking is observed and the junction current grows only in the vicinity of $\omega_0$. Above a critical drive power, $P_c$ in Fig.\ \ref{AR}, the resonator remains phase locked and its amplitude continues to grow up to a deterministic maximum, set either by damping (region 2) or by $I_0$ (region 3). The existence of a threshold for phase-locking, which obeys a universal scaling law, was first observed in the context of non-neutral plasmas \cite{Fajans99}; our work is the first observation of this transition in a micro-electronic circuit operating at GHz frequency and mK temperature\textemdash five orders of magnitude higher in frequency and lower in temperature than Ref.\ \onlinecite{Fajans99}. 

We proceed to briefly analyze the threshold phenomenon (a detailed analysis will be given elsewhere). The dynamics of a Josephson junction, ac biased through a resonant cavity near its resonance, can be modeled by the equivalent series $RLC$ circuit shown in Fig.\ \ref{circuit}b \cite{Vlad07,Boaknin07}, where $\tilde{L}\equiv L+L_\textrm{J}(0)\approx\pi Z_0/2\omega_0=7.74$ nH, $C\approx 2/\pi Z_0\omega_0=1.25$ pF, and $\omega_0^2=1/\tilde{L}C$. The effective loss due to external loading is $R_\textrm{eff}=\pi Z_0/2Q$, and the effective drive amplitude is $V_\textrm{d}=V_\textrm{in}\sqrt{\pi/Q}$ where $V_\textrm{in}=\sqrt{2Z_0P_\textrm{in}}$ is the amplitude of the incident wave referred to a matched load. From Kirchoff's voltage law we obtain an equation of motion for the charge $\tilde{q}(t)=\int I(t')dt'$ in the circuit. Defining the non-linearity ratio $2a=L_\textrm{J}(0)/\tilde{L}$, and the dimensionless time $\tau=\omega_0t$, charge $q=(\omega_0\sqrt{2a}/I_0)\tilde{q}$, and drive $\epsilon=(\sqrt{2a}/\tilde{L}I_0\omega_0)V_\textrm{d}$, and expanding the voltage drop across the junction to second order in $I/I_0$, the equation of motion in the weakly non-linear regime \cite{Fajans99} becomes:
\begin{equation}\label{dimless}
\ddot{q}(1+\dot{q}^2/2)+q+\nu\dot{q}=\epsilon\cos\phi_\textrm{d},
\end{equation}
where $\nu=1/Q$, $\phi_\textrm{d}=\omega_0t-\alpha t^2/2$ is the phase of the chirped drive, and the derivatives are with respect to $\tau$.

Transforming to a chirped frame rotating with the drive [where equation (\ref{dimless}) reduces to the Duffing model\cite{Vlad07}], and neglecting fast oscillating terms, we cast equation (\ref{dimless}) into an equation for the complex variable $\Psi=A\exp(i\Theta)$, where $A$ is proportional to the amplitude of $q(\tau)$, and $\Theta$ is the phase mismatch between $q(\tau)$ and the drive:
\begin{equation}\label{rotating}
i\frac{d\Psi}{d\tilde{\tau}}+\left(|\Psi|^2-\tilde{\tau}+i\gamma/2\right)\Psi=\mu.
\end{equation}
Here $\tilde{\tau}=\alpha^{1/2}\tau/\omega_0$, $\gamma=\omega_0\alpha^{-1/2}/Q$, and $\mu=\epsilon\:\omega_0^{3/2}\alpha^{-3/4}/8$. Equation (\ref{rotating}), with the initial condition $\Psi=0$ at $\tilde{\tau}\rightarrow-\infty$, admits two asymptotic solutions at $\tilde{\tau}\rightarrow+\infty$: a solution decoupled from the drive for $\mu<\mu_{cr}$, in which $\Theta$ grows quadratically with time, and a phase locked solution $\Theta=0$ for $\mu>\mu_{cr}$. 

In the absence of damping \cite{Friedland02} the transition between the two solutions of equation (\ref{rotating}) occurs for $\mu_{cr}^0=0.41$. When damping is present, $\mu_{cr}$ increases and becomes dependent on the damping: $\mu_{cr}(\gamma)\simeq\mu_{cr}^0(1+a\gamma+b\gamma^2)$ for $\gamma\ll 1$, with coefficients $a=1.06$ and $b=0.67$ found numerically. 

Combining the definitions above for $\mu$ and $\epsilon$, we find the critical drive amplitude for phase locking:
\begin{equation}\label{Vc}
V_\textrm{d}^{cr}=8\:\sqrt{\frac{2e}{\hbar\omega_0}}\:(\tilde{L}I_0)^\frac{3}{2}\alpha^\frac{3}{4}\mu_{cr}.
\end{equation}
The scaling law $V_\textrm{d}^{cr}\propto\alpha^{3/4}$ is exact only in a lossless resonator. For finite dissipation, $\gamma\propto\alpha^{-1/2}$ and corrections to this law enter through $\mu_{cr}(\gamma)$.
  
\begin{figure}
\epsfxsize=3.2in
\epsfbox{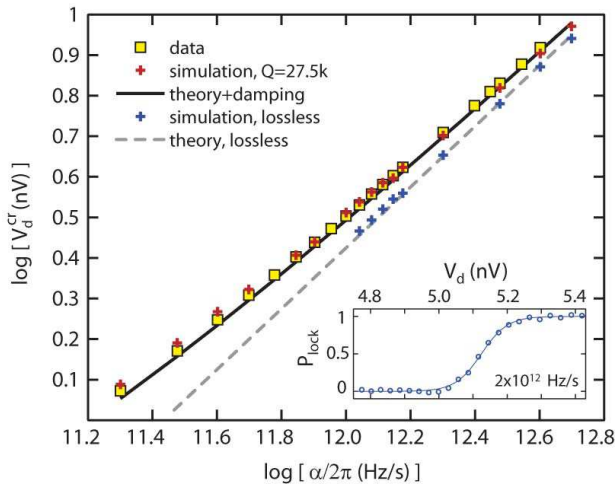}
\caption{\label{threshold} {\bf Critical drive scaling and phase-locking probability curve.} Comparison of the critical drive voltage $V_\textrm{d}$ obtained from the experiment ($\Box$) to equation (\ref{Vc}) with (solid line) and without (dashed line) dissipation. Also shown are values for the critical drive from simulations of the fully non-linear equation of motion for the equivalent $RLC$ circuit ($+$), for lossless resonator (blue) and for $Q=2.75\times10^4$ (red). Inset: probability of phase-locking for drive amplitudes near threshold at $\alpha/2\pi=2\times10^{12}$ Hz/s, evaluated over a 0.5 MHz frequency band centered at 1.6095 GHz. Solid line is a sigmoidal fit.}
\end{figure}

To compare the observed value of the phase-locking threshold, $V_\textrm{d}^{cr}$, to the theory outlined above, we first note that the probability of phase-locking, P$_\textrm{lock}$, obtained from averaging the response from 5,000 frequency sweeps \cite{avg_note}, grows from zero to one over a finite range of drive amplitudes. A typical phase-locking probability `s-curve' is plotted in the inset of Fig.\ \ref{threshold} as a function of $V_\textrm{d}$, for $\alpha/2\pi=2\times10^{12}$ Hz/s, where we define the threshold at P$_\textrm{lock}=0.5$. We measured $V_\textrm{d}^{cr}$ as a function of chirp rate, and the results are plotted in Fig.\ \ref{threshold} ($\Box$). We also performed numerical simulations of the fully non-linear equation of motion for $\tilde{q}(t)$ in the series $RLC$ model of Fig.\ \ref{circuit}b, without making any approximations to the voltage drop across the junction, and using experimentally determined parameters. We simulated both a resonator with $Q=2.75\times10^4$ and a lossless one. The numerical simulations ($+$) and the predictions from equation (\ref{Vc}) with (solid line) and without (dashed line) damping are also plotted in Fig.\ \ref{threshold}.  We observe excellent agreement between experiment and theory if damping is included. Note that this is not a fit\textemdash all parameters are fixed to their experimentally determined values. Moreover, the agreement of equation (\ref{Vc}) with the fully non-linear simulations confirms the validity of the weak non-linearity assumptions made in the threshold analysis\textemdash the phase-locking transition occurs at small amplitudes.

From the measured width of the phase-locking threshold we can estimate the potential sensitivity of the JCA, where as a benchmark we consider detecting a 1 \% variation in $I_{0}$, a typical signal associated with the transition between the ground and excited state of a superconducting `Quantronium' qubit \cite{Metcalfe07}. The discrimination power of the device can be estimated from the fractional change in critical current, $\Delta I_0/I_0$, that will shift the `s-curve' by an amount equal to the threshold width: $\Delta I_0/I_0=2\Delta V_\textrm{d}^{cr}/3V_\textrm{d}^{cr}$ from equation (\ref{Vc}). For the data in the inset of Fig.\ \ref{threshold}, and defining the width $\Delta V_\textrm{d}^{cr}$ for $0.27<\textrm{P}_\textrm{lock}<0.73$, we find $\Delta I_0/I_0=9.6\times10^{-3}$. The 1 \% variation in $I_{0}$ (6.1 nA for our device) can be resolved with $\sim46\;\%$ contrast; the ultimate sensitivity of the JCA requires an understanding of the dependence of the threshold width on noise, both classical and quantum, which is currently being pursued. For this chirp rate of $\alpha/2\pi=2\times10^{12}$ Hz/s, a single measurement can be accomplished in less than 10 $\mu$s. Significantly faster chirp rates and shorter measurement times are in principle possible.

The dynamics associated with operating the amplifier with a frequency modulated (FM) drive versus an amplitude modulated (AM) drive are quite different. In Fig.\ \ref{phase}b,d we show the calculated in-phase and quadrature phase components of $I/I_0$ for AM and FM drives, respectively. In the AM case, for detuning $\omega<\omega_0$, the system can switch from a low-amplitude oscillation attractor (A) to a high-amplitude oscillation one (B), with significant phase oscillations associated with the switching as shown in Fig.\ \ref{phase}b. In fact, for large enough detuning, the junction's phase can exceed $\delta=\pi$, which can induce state mixing when measuring a quantum system. Trajectories for the chirped drive are shown in Fig.\ \ref{phase}d, where the two traces correspond to the experimental data shown in Fig.\ \ref{phase}c. For sub-critical drives (blue), the system builds amplitude as the chirp frequency crosses $\omega_0$, but remains in the low-oscillation state and eventually relaxing back to zero amplitude. For super-critical drive (red), the system smoothly follows the evolution of the high-amplitude attractor without spurious oscillations of the phase. Another feature of the JCA is that we can potentially perform quantum measurement with very few photons (near $\omega_0$), but ``latch'' and record the signal with a large number of photons at a final frequency $\omega<\omega_0$, so that minimal fidelity is lost due to noise in the measurement electronics.                

In summary, we have observed that a high-$Q$ resonator embedding a Josephson junction subject to a chirped drive exhibits a transition to phase-locking with a sharp threshold that can be used for detecting small changes in $I_0$. This device\textemdash the Josephson chirped amplifier\textemdash provides a new route for non-linear, dispersive quantum measurement and a test bed for non-equilibrium quantum statistical mechanics in a chirped reference frame. We studied the dependence of the critical drive on the chirp rate and found excellent agreement with both theory and simulations.  The phase-locked highly excited state and the associated phase-locking threshold are a robust feature of chirped non-linear oscillators, and should be observable in any system with low loss and weak non-linearity: electrical, mechanical, or photonic, for example. Finally, we note that another approach to observing these phase-locking effects is to modulate the resonant frequency of the oscillator in the presence of a harmonic drive; a candidate technology for such tuning was recently demonstrated in Ref.\ \onlinecite{Sandberg08} in the context of variable coupling elements for qubits.

The authors thank R. Vijayaraghavan, V. Manucharyan, and J. Clarke for useful discussions. Financial support was provided by the Office of Naval Research under grant N00014-07-1-0774 (O.N., I.S.), UC Berkeley Chancellor's Faculty Partnership Fund (I.S.), the Hellman Family Faculty Fund (I.S.), the US-Israel Binational Science Foundation under grant No. 2004033 (L.F., J.W.), and the US Department of Energy under grant No. DE-FG02-04ER41289 (J.W.).


\begin{thebibliography}{10}
\bibitem{avg_note}
\bibinfo{note}{Large number of averages is needed to observe the threshold
  behavior at low power, with a high bandwidth and a system noise temperature
  of $T_N\sim9$ K}.
\end{thebibliography}

\end{document}